\pdfoutput=1

\documentclass[11pt,twoside,a4paper,cmspaper,final,collab]{cms-tdr}
\def\svnVersion{494657P}\def\cmsCernNoTag{CERN-EP-2018-325}\def\cmsCernDate{\today}\def\cmsMessage{Published in the Journal of High Energy Physics as \href{http://dx.doi.org/10.1007/JHEP05(2019)043}{\doi{10.1007/JHEP05(2019)043.}}}

\begin{document}\cmsNoteHeader{FSQ-17-001}

\hyphenation{had-ron-i-za-tion}
\hyphenation{cal-or-i-me-ter}
\hyphenation{de-vices}
\RCS$HeadURL: svn+ssh://mvandekl@svn.cern.ch/reps/tdr2/papers/FSQ-17-001/trunk/FSQ-17-001.tex $
\RCS$Id: FSQ-17-001.tex 465636 2018-06-22 09:09:16Z mvandekl $

\newlength\cmsFigWidth
\ifthenelse{\boolean{cms@external}}{\setlength\cmsFigWidth{0.85\columnwidth}}{\setlength\cmsFigWidth{0.4\textwidth}}
\ifthenelse{\boolean{cms@external}}{\providecommand{\cmsLeft}{top\xspace}}{\providecommand{\cmsLeft}{left\xspace}}
\ifthenelse{\boolean{cms@external}}{\providecommand{\cmsRight}{bottom\xspace}}{\providecommand{\cmsRight}{right\xspace}}
\cmsNoteHeader{FSQ-17-001}

\title{Measurement of inclusive very forward jet cross sections in proton-lead collisions at $\sqrtsNN=5.02\TeV$}
\date{\today}

\newcommand{\QGSJet} {{\textsc{qgsjetii-04}}\xspace}
\newcommand{\EPOS} {{\textsc{epos-lhc}}\xspace}
\newcommand{\ROOUNFOLD} {{\textsc{roounfold}}\xspace}
\newcommand{\katie} {{\textsc{katie}}\xspace}
\newcommand{\Cascade} {{\textsc{cascade}}\xspace}
\newcommand{\AAMQS}{{\textsc{aamqs}}\xspace}
\newcommand\T{\rule{0pt}{2.6ex}}
\newcommand\B{\rule[-1.2ex]{0pt}{0pt}}

\newlength\cmsTabSkip\setlength{\cmsTabSkip}{1ex}

\abstract{
Measurements of differential cross sections for inclusive very forward jet production in proton-lead collisions as a function of jet energy are presented.  The data were collected with the CMS experiment at the LHC in the laboratory pseudorapidity range $-6.6<\eta<-5.2$. Asymmetric beam energies of 4\TeV for protons and 1.58\TeV per nucleon for Pb nuclei were used, corresponding to a center-of-mass energy per nucleon pair of $\sqrtsNN=5.02\TeV$.
Collisions with either the proton (\Pp+Pb) or the ion (Pb+\Pp) traveling towards the negative $\eta$ hemisphere are studied. The jet cross sections are unfolded to stable-particle level cross sections with $\pt\gtrsim3\GeV$, and compared to predictions from various Monte Carlo event generators.
In addition, the cross section ratio of \Pp+Pb and Pb+\Pp\ data is presented. The results are discussed in terms of the saturation of gluon densities at low fractional parton momenta. None of the models under consideration describes all the data over the full jet-energy range and for all beam configurations. Discrepancies between the differential cross sections in data and model predictions of more than two orders of magnitude are observed.}

\hypersetup{
pdfauthor={CMS Collaboration},
pdftitle={Measurement of inclusive very forward jet cross sections in proton-lead collisions at sqrt(s[NN])=5.02 TeV},
pdfsubject={CMS},
pdfkeywords={CMS, physics, jets, heavy ions}}

\maketitle

\newpage
\section{Introduction}
\label{sec:introduction}
In hadron-hadron collisions at high energies, scatterings with large momentum transfer
among the constituent partons lead to the production of jets. Such processes are frequently described by perturbative  quantum chromodynamics (QCD) calculations in the framework of collinear factorization, in which hard partonic cross sections are convolved with parton distribution functions (PDFs).

Jet production in hadron-hadron interactions is thereby
a useful tool to study the parton structure of hadrons. The so-called
Dokshitzer--Gribov--Lipatov--Altarelli--Parisi (DGLAP) equations~\cite{DGLAP:Altarelli,DGLAP:Dokshitzer,DGLAP:Gribov}
describe the evolution of the parton densities as a function of the virtuality $Q^2$
(\ie, the squared four-momentum exchange) of the hard scattering process. This approach has been
shown to be overall successful for the description of the production of high-\pt jets in proton-proton
collisions at the LHC~\cite{Khachatryan:2011zj,Chatrchyan:2011qta,Chatrchyan:2012gwa,Chatrchyan:2012pb,Khachatryan:2016hkr,Sirunyan:2017skj,Aad:2010ad,Aad:2011jz,Aad:2011fc,Aad:2013tea,Aad:2014pua,Aad:2015xis}.

At leading order (LO) in the QCD coupling  \alpS,
the pseudorapidity, $\eta$, and the \pt of a jet are related to the fraction $x$ of the momentum of the parent nucleon that is carried by
the incoming parton via
\begin{linenomath*}
\begin{equation}
x \approx\frac{\pt}{\sqrt{s}} {\re}^{\pm\eta},
\end{equation}
\end{linenomath*}
where $\sqrt{s}$ is the center-of-mass energy of the hadron-hadron collision. Forward jets with low \pt
therefore probe the parton densities and their evolution at small (and large) $x$. The measurements presented
in this paper, for jets with $\pt\geq3\GeV$ at very forward pseudorapidities $-6.6 < \eta < -5.2$,
are thereby sensitive to fractional momenta down to $x\simeq10^{-6}$.
Collinear factorization and DGLAP have been shown to be successful for the description of processes involving large momentum exchanges and moderate fractional momenta carried by the interacting partons. However, in the low-$x$ kinematic regime considered in this paper, this approach is expected to fail, whereas
the Balitsky--Fadin--Kuraev--Lipatov (BFKL) equations~\cite{BFKLGluonDensity1,BFKLGluonDensity2,BFKLGluonDensity3},
which evolve the parton densities as a function of $1/x$, should be better suited to describe forward jet production.

Both the DGLAP and BFKL equations are linear equations, \ie, they account for parton splitting and radiation, but not for parton recombination, processes. The BFKL and DGLAP equations predict a rapid rise of the gluon density towards small $x$, a result that has been experimentally confirmed by measurements at HERA~\cite{HeraF2Measurement}. This rise is mitigated
when next-to-leading-order (NLO) corrections are taken into consideration. Despite this, the growth
of the gluon density with decreasing $x$ ultimately will result in a violation of unitarity (\ie, the cross section
for parton scatterings will exceed the total inelastic hadronic cross section), and the linear evolution
equations alone will not be sufficient to describe forward jet data. To solve this, it has been hypothesized~\cite{DGLAP:Gribov}
that at sufficiently small values of $x$, nonlinear gluon recombination processes will slow-down the
uncontrolled growth of the PDFs. Such a parton ``saturation'' regime, characterized by a virtuality scale
known as the saturation scale $Q_{\mathrm{s}}(x)$, is described by the Balitsky--Kovchegov (BK)
evolution equations~\cite{BKGluonDensity1,BKGluonDensity2}. Saturation effects are expected to become important
in the kinematic region where the gluon density, $xg(x,Q^2)$, times the transverse extent of the gluons, $\alpS(Q^2)/Q^2$,
becomes equal to the transverse area of the hadron $\pi R_{\text{had}}^2$ \cite{InelasticpARef26}:
\begin{linenomath*}
\begin{equation}
\frac{\alpS(Q_{\mathrm{s}}^2)}{Q_{\mathrm{s}}^2} xg(x, Q_{\mathrm{s}}^2) \approx \pi R_{\text{had}}^2.
\end{equation}
\end{linenomath*}
In a heavy ion with number of nucleons $A$, the squared saturation scale $Q_s^2$ is expected to increase by a factor of $A^{1/3}$ with respect to that of a single nucleon, namely by approximately a factor of six for a lead nucleus.
For $x\approx10^{-6}$, the anticipated saturation scale in a lead nucleus is approximately $10\GeV^2$ and, thereby, enhanced
signals of gluon saturation are expected when colliding lead ions compared to protons at the LHC.
Besides being a major research topic in its own right, parton saturation is also an important theoretical ingredient for describing
the initial state of heavy ion collisions~\cite{Albacete:2012xq}, and for understanding cosmic ray data~\cite{Albacete:2016tjq}.

Various measurements of jet production in proton-lead collisions at $\sqrtsNN=5.02\TeV$ at the CERN LHC have already been
performed~\cite{Adam:2015hoa,ATLAS:2014cpa,Khachatryan:2016xdg}.
In addition, the ATLAS and CMS experiments have reported forward jet measurements in proton-proton collisions at
$\sqrt{s}=7\TeV$ for \pt values above 20\GeV and within $\abs{\eta}<5.2$~\cite{Chatrchyan:2012gwa,Aad:2015xis}.
Dijet measurements in proton-proton collisions have also been performed by ATLAS~\cite{Aad:2014pua} and CMS~\cite{Khachatryan:2016udy},
in which the sensitivity to BFKL effects is enhanced by studying events that contain jet pairs with large rapidity separations.
In all these studies, some results favored BFKL-based approaches, while others were better described by models based on DGLAP evolution.

The measurement presented here exploits the CASTOR calorimeter~\cite{testbeamdata} on one side of the CMS experiment, which extends the
range in which jets can be measured down to $\eta=-6.6$, with a lower limit on the accessible jet \pt of approximately
3\GeV. Because of its unrivaled acceptance, a study of jets using CASTOR in proton-lead collisions has a unique sensitivity to the
low-$x$ regime and to perturbative nonlinear parton evolution effects.

In this paper, differential cross sections for inclusive forward jet production as a function of jet energy in hadronic,
nondiffractive proton-lead collisions at $\sqrtsNN=5.02\TeV$ are presented. Collisions in which either the proton (\Pp+Pb) or
the $^{208}\text{Pb}$ ion (Pb+\Pp) travels towards the negative $\eta$ hemisphere (where CASTOR is located) are studied.
Asymmetric beam energies of 4\TeV for the protons and 1.58\TeV per nucleon for the lead nuclei were used, yielding a proton-nucleon center-of-mass energy of $\sqrtsNN=5.02\TeV$.
This implies that the center-of-mass system of the \Pp+Pb collisions is shifted with respect to the laboratory frame, and the Pb+\Pp~center-of-mass system is boosted in the opposite direction. Therefore, the effective acceptance of CASTOR in the center-of-mass frame is different for the \Pp+Pb and Pb+\Pp\ beam setups. Jet cross sections, unfolded to the stable-particle level, and ratios of \Pp+Pb to Pb+\Pp{} cross sections are studied,
and compared to the predictions from Monte Carlo (MC) event generators including, or not, parton saturation effects.

This paper is organized as follows. In Section~\ref{sec:detector} the CMS detector is presented.  Section \ref{sec:generators}
discusses the various models that are used in this study. Section~\ref{sec:analysis} presents the experimental analysis,
followed by an overview of systematic uncertainties in Section~\ref{sec:systematics}.
A comparison of the results to MC predictions is given in Section~\ref{sec:results}, followed by the summary in Section~\ref{sec:summary}.

\section{The CMS detector}
\label{sec:detector}

The central feature of the CMS apparatus is a superconducting solenoid of 6\unit{m} internal diameter, providing a magnetic field of 3.8\unit{T}. Within the solenoid volume are a silicon pixel and strip tracker, a lead tungstate crystal electromagnetic calorimeter (ECAL), and a brass and scintillator hadron calorimeter (HCAL), each composed of a barrel and two endcap sections. Forward calorimeters extend the $\eta$ coverage provided by the barrel and endcap detectors. Muons are detected in gas-ionization chambers embedded in the steel flux-return yoke outside the solenoid.

The forward hadron (HF) calorimeter uses steel as an absorber and quartz fibers as the sensitive material. The two HF detectors are located at 11.2\unit{m} from the interaction point, one on each side of CMS, and together they provide coverage in the range $3.0<\abs{\eta}<5.2$.

\begin{figure}[hbtp]
\centering
\includegraphics[width=0.55\textwidth]{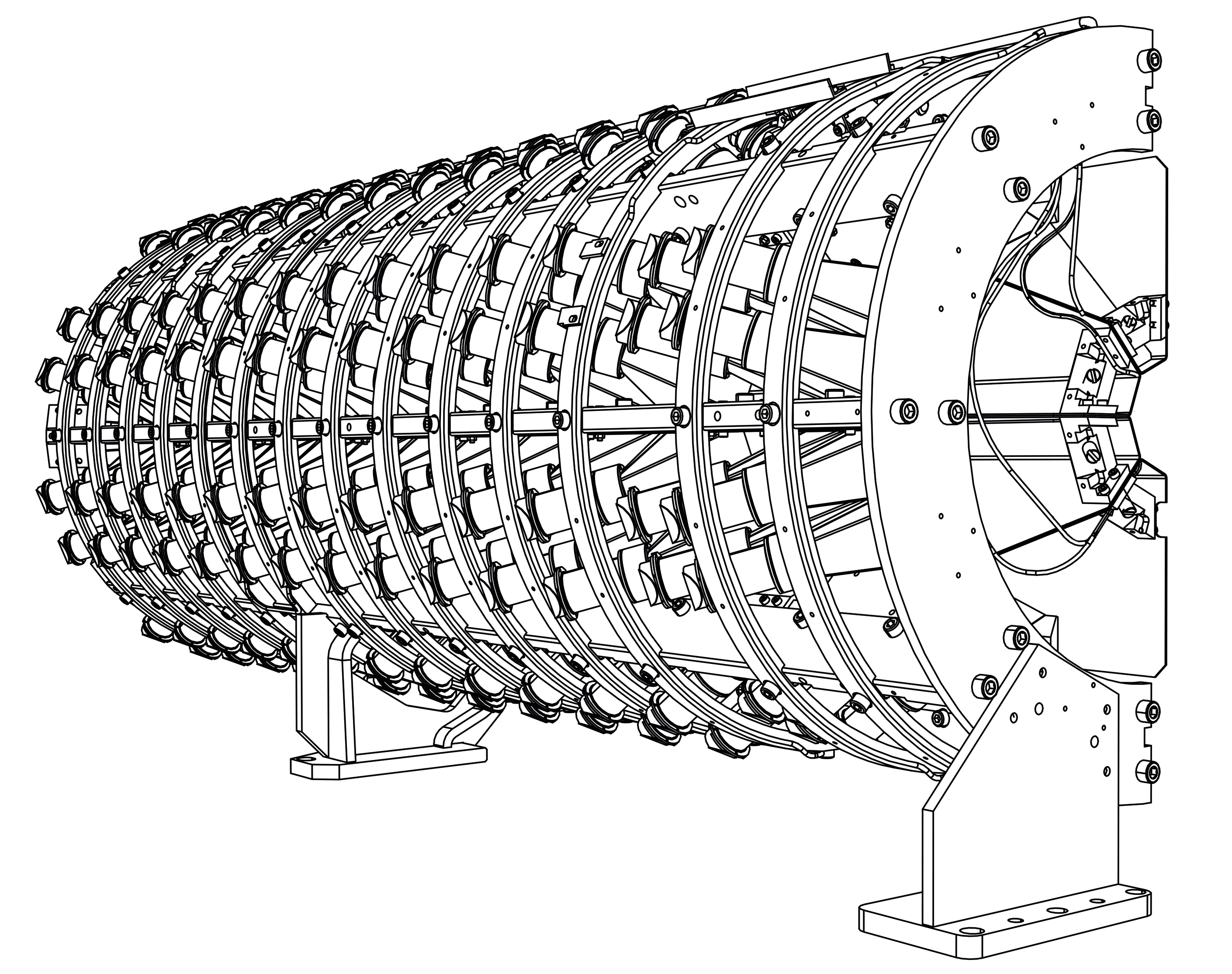}
\caption{A schematic drawing of one half of the CASTOR calorimeter and its mechanical support structure. The diameter of CASTOR is roughly 0.6\unit{m} and it is approximately 1.6\unit{m} in length. The transversal and longitudinal segmentation in eight sectors and fourteen modules, respectively, can be clearly distinguished. The 112 small cylinders represent the photomultiplier tubes of CASTOR. These are mounted on light guides, which transport the Cherenkov radiation out of the detector. It may be observed that CASTOR has only transverse and no $\eta$ segmentation.}
\label{CASTORFIG}
\end{figure}

The very forward angles, \ie, $-6.6<\eta<-5.2$, are covered at one end of CMS by the CASTOR calorimeter, located at 14\unit{m} from the interaction point. A diagram of CASTOR is shown in Fig.~\ref{CASTORFIG}. It consists of tungsten absorbers and quartz detection plates. The CASTOR calorimeter is segmented into 16 azimuthal sectors and each sector is longitudinally segmented into 14 modules. The first two modules comprise the electromagnetic section, which is approximately 20 radiation lengths deep. The remaining twelve modules constitute the hadronic section and are approximately nine nuclear interaction lengths deep. Thus, the overall depth of CASTOR is approximately ten nuclear interaction lengths. CASTOR is a non-compensating calorimeter with a relative response to incident pions with respect to electrons of approximately 50\% at an energy of 100\GeV.
The intrinsic relative jet-energy resolution of CASTOR is approximately 25 (10)\% for fully contained jets (\ie, jets that do not suffer from transverse leakage) of 550 (2500)\GeV. It should be noted though that CASTOR has no $\eta$ segmentation; this leads to substantial transverse leakage effects. Consequently, the \textit{effective} relative jet-energy resolution deteriorates to approximately 60 (25)\% at  550 (2500)\GeV.

A more detailed description of the CMS detector, together with a definition of the coordinate system used and the relevant kinematic variables, can be found in Ref.~\cite{CMSTDR}.

\section{Event generators}
\label{sec:generators}

Monte Carlo event generators are used to correct the data for acceptance and instrumental effects, as well as to compare to the experimental results.  Events, for which the detector response is fully simulated, are generated with the \EPOS~\cite{EPOS-LHC}, \HIJING v1.383~\cite{Hijing}, and \QGSJet~\cite{QGSJet2}  programs.   The response of the CMS detector is simulated using \GEANTfour~\cite{Geant4}, including the CASTOR calorimeter and its energy response, which was validated using test-beam measurements~\cite{testbeamdata}.  Two further Monte Carlo models, \katie (version \katie-2017-05-09)~\cite{KatieSoftware} and \AAMQS~\cite{Albacete:2016tjq}, are used to compare model predictions to the unfolded measurements.  None of these generators include photon-induced events. In addition, the \HIJING, \katie, and \AAMQS  generators do not include diffractive events.

The \HIJING event generator models hard QCD interactions using the \PYTHIA v5.7~\cite{Pythia} event generator, based on leading order collinear factorization and parton showering with DGLAP evolution. The Eichten-Hinchliffe-Lane-Quigg parameterization of the nucleon pdf is used~\cite{RevModPhys.56.579}. The soft interactions are modeled with the Lund \textsc{fritiof}~\cite{FRITIOF} and dual-parton~\cite{DualParton} models. In addition, \HIJING incorporates saturation effects via nuclear shadowing~\cite{InelasticpARef26, Hijingref49,Hijingref50}.  These shadowing corrections are obtained from a fit of the ratio of nucleus to proton sea quark structure functions at moderate $x$ values ($x>10^{-3}$), which are then extrapolated to lower $x$.

The \EPOS program is an update of \textsc{epos} version 1.99 that has been specifically tuned to reproduce the first LHC \Pp+\Pp, \Pp+Pb, and Pb+Pb measurements. This model uses a combination of soft pomeron exchange (as in Regge--Gribov theory~\cite{Gribov:1968fc,Gribov:1968uy}) and a semihard contribution based on the convolution of a nonperturbative pre-evolution, a DGLAP-based hard evolution, and standard leading order QCD $2 \to 2$ cross sections~\cite{Drescher:2000ha}.  The semihard contribution has been tuned to HERA structure function data and is recast as a hard pomeron amplitude.  Saturation is modeled through pomeron-pomeron interactions and is implemented by modifying the $x$ dependence of the pomeron amplitudes. In Ref.~\cite{Pierog:2015yzd}, however, it is shown that this leads to too strong a suppression of the hard component, a shortcoming that will be remedied in an upcoming version of the model.

Similar to \EPOS, the \QGSJet model is based on the Regge--Gribov theory for the soft interactions and hard matrix elements convolved with DGLAP evolution for the hard scatterings. In this case, pomeron self-interactions result in saturation effects.
The Cosmic Ray Monte Carlo \textsc{CRMC} package v1.5.6~\cite{CRMC} is used to generate the cross sections for the \EPOS, \HIJING, and \QGSJet models.

The \katie program allows calculations of tree-level matrix elements with offshell initial-state partons.  It is therefore ideally suited to be used in combination with transversal-momentum dependent PDFs,  known as ``TMDs".

In this paper, \katie is used in a hybrid high-energy factorization approach where high-$x$ partons in one incoming hadron are treated in the collinear framework, while low-$x$ partons in the other incoming hadron are modeled using a TMD \cite{Dumitru:2005gt, Stasto:2016wrf, HEFpaper1, HEFpaper2}.  Such an approach is considered to be valid in configurations with very asymmetric fractional momenta, resulting, for example, in forward jet production.  The TMDs are obtained from the Kutak--Sapeta (KS) framework \cite{Kutak:2012rf}, combined with linear (BFKL) or nonlinear (BK) evolution with running coupling, respectively.  The strength of the nonlinear term can be varied; here, a default strength of 75\% is used, with a variation of $\pm25\%$ to reflect the sensitivity to nonlinear evolution.  The parton density used in the collinear framework is the CTEQ10 NLO set~\cite{Lai:2010vv}. The \katie program is used to calculate leading order hard scattering matrix elements for the $2 \to 1$ process ($\Pg^\ast \Pq \to \Pq$ or $\Pg^\ast \Pg \to \Pg$) with incoming offshell gluons. The subsequent hadronization is modeled using \Cascade v2.4.13~\cite{Cascade}. A more in-depth description of the predictions that are included in the present paper can be found in Ref.~\cite{pAPredictions_KVV}.

The \AAMQS predictions~\cite{Albacete:2016tjq} for the forward jet spectra are also obtained using hybrid factorization. At the parton level, the \AAMQS model generates quarks and gluons from $\Pg \Pq \to \Pq$ and $\Pg\Pg \to \Pg$ hard processes, along with initial- and final-state radiation based on DGLAP evolution.  The TMD is related to a dipole scattering amplitude in coordinate space via a Fourier transform and, in this paper, an implementation based on the McLerran--Venugopalan (MV) model~\cite{McLerran:1993ni,McLerran:1993ka,McLerran:1994vd} is used to derive the dipole scattering amplitude within the color glass condensate framework~\cite{BKGluonDensity1,JalilianMarian:1997jx,JalilianMarian:1997dw,BKGluonDensity2,Weigert:2000gi,Iancu:2000hn,Ferreiro:2001qy,Gelis:2010nm}. In this framework the hard partons act as color sources for a classical non-Abelian background field, which is formed by numerous soft gluons.
The \AAMQS calculations are performed at leading order in \alpS. The evolution of the TMD is performed using the BK equation with running coupling strength.  For the collinear PDFs the CTEQ6 LO set was used~\cite{Pumplin:2002vw}.  Multiple parton interactions are included in the framework, and hadronization is performed using the Lund string fragmentation model~\cite{ANDERSSON198331}, as implemented in \PYTHIA. To account for missing higher-order effects in the \katie and \AAMQS models, the overall jet cross section normalization may be modified using a constant $K$-factor. This factor is fixed to unity for the predictions presented in this paper. Full next-to-leading order calculations~\cite{Ducloue:2017dit} may lower the jet cross sections by approximately 40\%.

The \katie and \AAMQS models do not provide any modeling of the ion remnants that did not directly participate in the hard interaction, but are nonetheless incident on CASTOR. These models are therefore not expected to be successful in describing interactions in which the incoming ion travels towards CASTOR (the so-called Pb+\Pp\ configuration).

\section{Data analysis}
\label{sec:analysis}
The \Pp+Pb and Pb+\Pp\ collisions that are analyzed in this study were delivered to CMS in 2013. Only data collected when the whole CMS detector was optimally functioning are used, corresponding to a total integrated luminosity of the \Pp+Pb and Pb+\Pp\ data sets of 3.13 and 6.71\nbinv, respectively. The peak pileup (\ie, the maximum average number of simultaneous interactions per bunch crossing) in these runs was 0.0763.

Events were collected using a minimum bias trigger~\cite{Khachatryan:2016bia} requiring the presence of a track in the pixel detector (with acceptance $\abs{\eta}\leq2.5$) with a \pt above 0.4\GeV. In addition, an energy deposit above 4\GeV in the HF calorimeters (with acceptance $3\leq\abs{\eta}\leq5$) on both sides is required offline. It has been shown in Ref.~\cite{PAS:InelasticSigma} that this selection effectively suppresses the contribution of photon-induced and diffractive events to a negligible level. To mitigate the effect of pileup in data, only events with at most one reconstructed primary vertex~\cite{Chatrchyan:2014fea} are selected. After the event selection has been applied, the CASTOR jet spectrum is found to be independent of the instantaneous luminosity of the collisions, confirming that pileup does not impact the results of this analysis.
 Dedicated event-selection filters are applied to remove beam-gas scattering events and interactions of the beam with the beam pipe.  By studying the occurrence of events satisfying the online and offline criteria in a zero bias event sample, in which only the presence of beams within the CMS detector is required, the trigger is found to be fully efficient for events that fulfill the offline selection criteria.

The reconstruction of jets in CASTOR is detailed in Ref.~\cite{CMS-DP-2014-022}.
The CASTOR calorimeter can be subdivided into towers, where a tower consists of all modules within a given longitudinal CASTOR sector.  These towers are clustered into jets using the anti-\kt algorithm~\cite{antiktalgo,Cacciari:2011ma,Cacciari:2005hq} with a distance parameter of 0.5. The distance parameter was optimized by studying the correlation between detector-level and particle-level jets, and between particle-level and parton-level jets. Because of the lack of $\eta$ segmentation in CASTOR, both towers and detector-level jets have their $\eta$ coordinate fixed to the geometrical center of the sector (\ie, $\eta=-5.9$).

The reconstructed jets in CASTOR require an energy-dependent correction factor because of the noncompensating nature of CASTOR. The jets are therefore first identified to be of hadronic origin, using topological shower properties such as depth, width and electromagnetic energy fractions, before the required calibration functions are applied.   These energy-dependent calibration functions are obtained from simulation as follows. First, a particle-level jet is matched to a detector-level jet. Subsequently, it is required that the particle-level jet is isolated (\ie, that there is no other particle-level jet contributing to the detector-level jet) and contained (\ie, the jet is incident in the region $-6.1<\eta<-5.7$). For these jet pairs the response is calculated. The inverse of the response function is applied to the reconstructed jets in order to correct for the noncompensating nature of the calorimeter. Details of the jet calibration and reconstruction are presented in Ref.~\cite{ThesisKlundert}.

\begin{figure}[hbtp]
\centering
\includegraphics[width=0.49\textwidth]{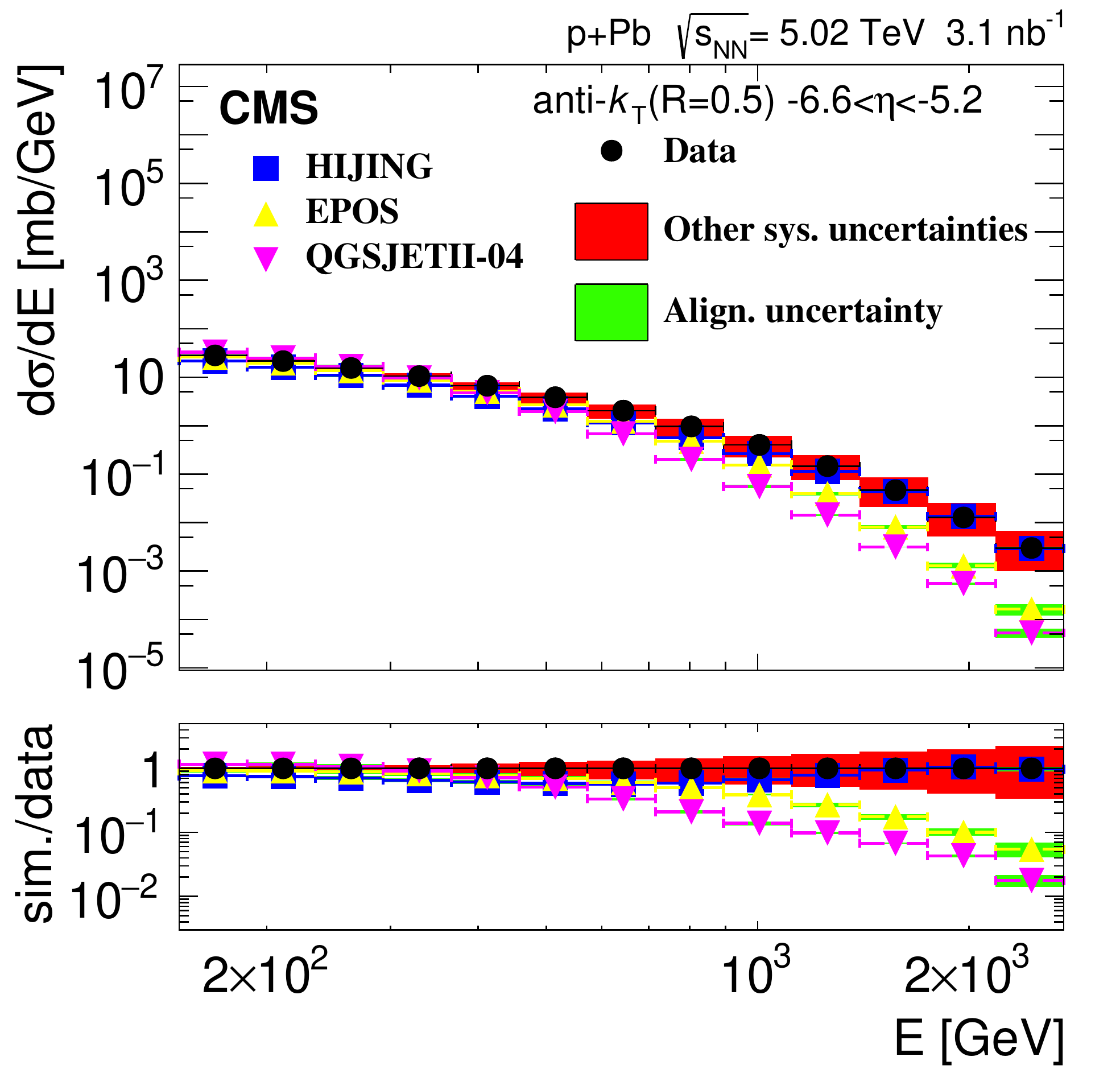}
\includegraphics[width=0.49\textwidth]{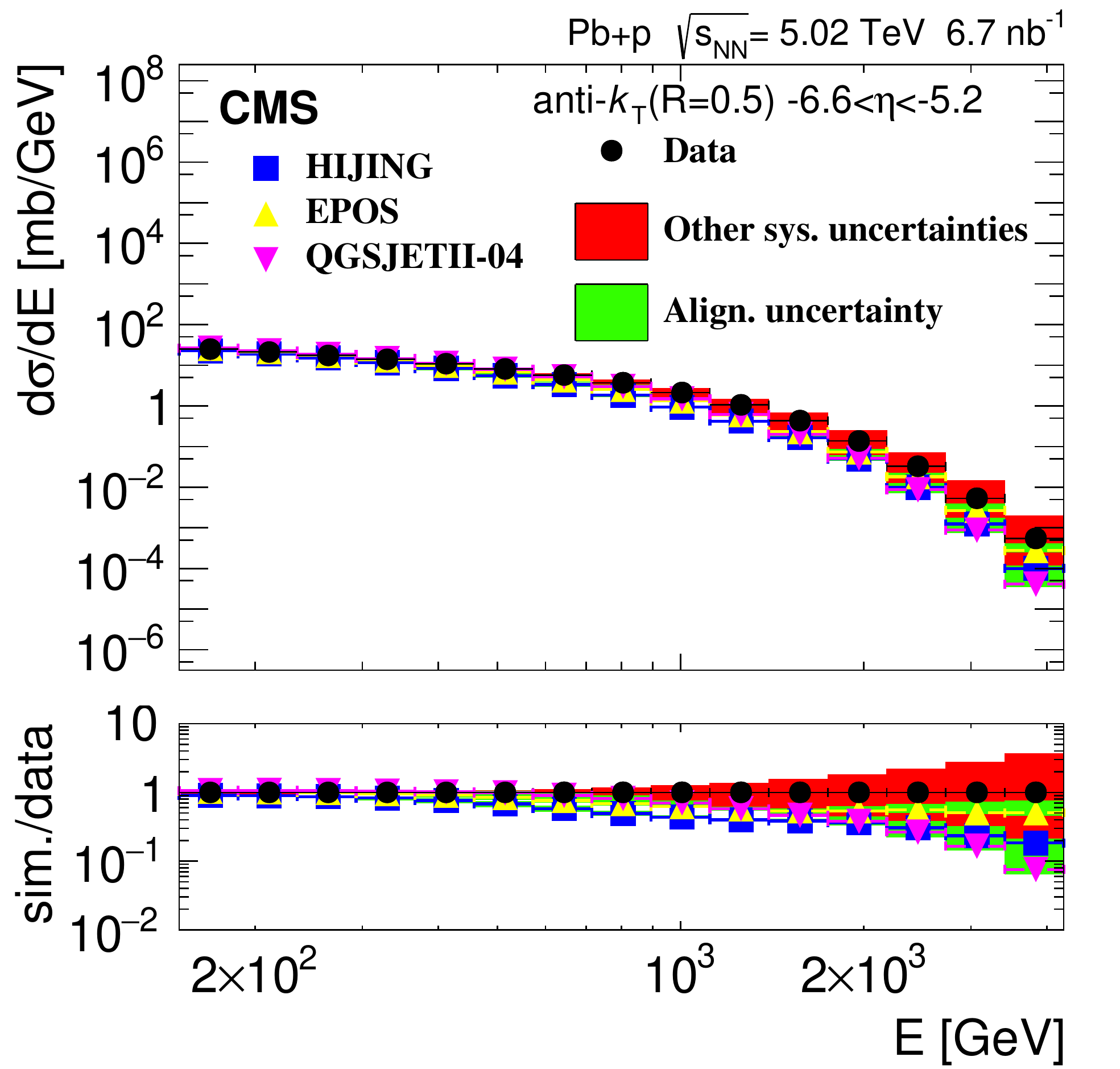}
\caption{Detector-level differential cross sections for inclusive forward jet production as a function of calibrated jet energy in \Pp+Pb (left) and Pb+\Pp\ (right) collisions. Model predictions are shown for \EPOS, \HIJING, and \QGSJet.}
\label{plot:RecoLevelJetESpectra}
\end{figure}

\begin{figure}[hbtp]
\centering
\includegraphics[width=0.49\textwidth]{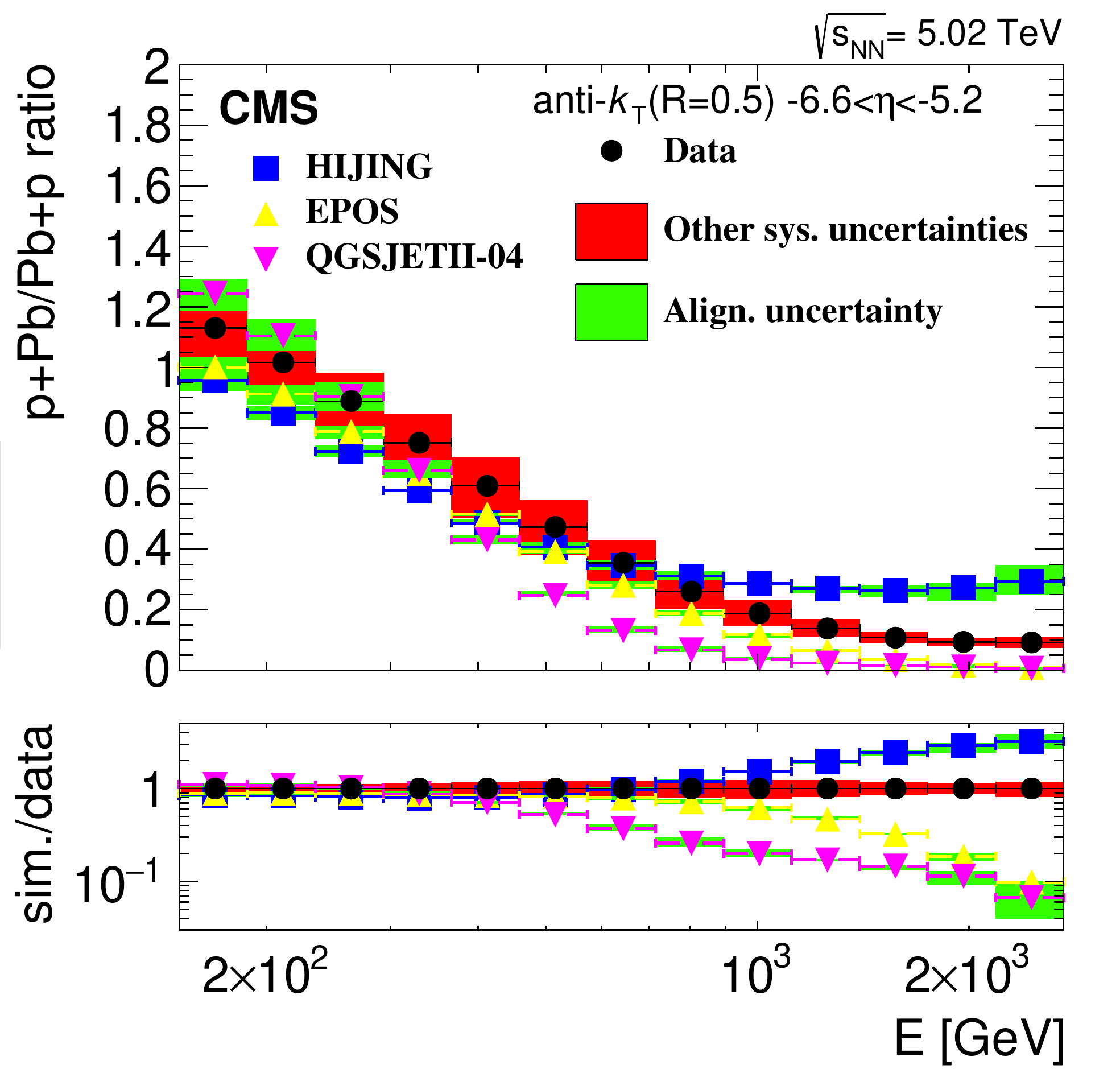}
\caption{Detector-level ratio of differential cross sections for inclusive forward jet production in \Pp+Pb to Pb+\Pp\ data vs.\@ calibrated jet energy.
Model predictions are shown for \EPOS, \HIJING, and \QGSJet.} \label{plot:RecoLevelJetESpectrapAAp}
\end{figure}

Detector-level differential cross sections as a function of calibrated jet energy $(E)$ are shown in Fig.~\ref{plot:RecoLevelJetESpectra} for $E>150\GeV$. The systematic uncertainty band on the data includes the energy scale, jet identification, and luminosity uncertainties, while the alignment uncertainty has been displayed for each model individually. These systematic uncertainties are described in Section~\ref{sec:systematics}. For the Pb+\Pp\ data and simulations, the energy spectrum extends well beyond the incoming energy per nucleon of 1.58 \TeV, because the jet algorithm picks up energy deposits originating from the ion remnants, for which the spectra are not corrected. The detector-level distributions indicate that \HIJING yields the best overall description of the data, although the description of the beam remnants can certainly be further improved. The \EPOS and \QGSJet models underestimate the measurements, in particular in the high-energy part of the \Pp+Pb spectrum. The ratio of energy spectra of \Pp+Pb to Pb+\Pp\ data is shown in Fig.~\ref{plot:RecoLevelJetESpectrapAAp}.

The data are corrected for detector efficiency and acceptance, bin migrations, and detector-level jets nonexistent at stable-particle level (fake jets) through an unfolding procedure.
In simulation, stable particles are defined to be those with a proper lifetime $c\tau>1\unit{cm}$; events at stable-particle level are required to have at least one particle in each $\eta$ hemisphere in the range $3<\abs{\eta}<5$ with a minimum energy of 4\GeV, and a charged particle  with $\abs{\eta}<2.5$ and $\pt>0.4\GeV$, thus closely following the
detector-level event selection. Jets are formed from the stable particles, with the exception of neutrinos, including their full four-momentum in the laboratory system, by using the anti-\kt jet algorithm with distance parameter 0.5, and are required to have their axis within the CASTOR acceptance range of $-6.6<\eta<-5.2$.

The unfolding of the data is performed using the D'Agostini iterative procedure~\cite{BayseianUnfoldingAgostini} as implemented in \ROOUNFOLD~\cite{roounfold}. To perform the unfolding, jets at the detector and stable-particle levels are matched in $\phi$ to create a response matrix. Because of the lack of $\eta$ resolution of CASTOR, it is not possible to match jet pairs in the distance $R$ between the detector- and particle-level jets. The procedure also corrects for stable-particle level jets not found at detector-level (missed jets) and for fake jets. It is found that 100 (\Pp+Pb) and 720 (Pb+\Pp) Bayesian iterations are needed to obtain a stable result. The relatively large number of iterations can be explained by the broadness of the response matrices. Two main effects are responsible for this broadness. One is the intrinsic energy resolution of CASTOR. The other is the lack of $\eta$ segmentation, which causes large transversal leakage. This leads to a further broadening of the response matrix. This is outlined in more detail in Ref.~\cite{ThesisKlundert}. In the unfolding procedure all jets with energy above 150\GeV are taken into account. However, in order to avoid large model-dependent correction factors arising from migrations of jets over the 150\GeV cutoff, the unfolded jet energy spectra are presented in Section~\ref{sec:results} starting from an energy threshold of 550\GeV, corresponding to a transverse momentum of $\pt=E/\cosh(\eta)\approx3\GeV$. From this cutoff onwards the fraction of missed and fake jets is smaller than 10 and 5\%, respectively. The unfolding procedure and its validation are discussed in Ref.~\cite{ThesisKlundert}. Finally, the cross section ratio of \Pp+Pb to Pb+\Pp\ is obtained from the individually unfolded spectra.

\section{Systematic uncertainties}
\label{sec:systematics}

The following sources of systematic uncertainty are investigated, given in order of the size of their contribution:

\begin{itemize}
\item \textit{Energy scale uncertainty.} The CASTOR energy scale uncertainty is $\pm 15\%$~\cite{forwardenergyflow13tev}. Therefore, the CASTOR jet-energy spectra are obtained with the energy scale varied in the reconstructed data by $\pm 15\%$ to estimate the associated uncertainty. The propagated energy scale uncertainty is largest in the \Pp+Pb spectrum, since this is the steepest descending spectrum.
\item \textit{Model dependence.} Different MC event generators (\EPOS, \HIJING, and \QGSJet) are each used to unfold the data. As shown in Fig.~\ref{plot:RecoLevelJetESpectra}, \HIJING is found to describe the data best and therefore this generator is used for the default unfolding procedure, while \EPOS and \QGSJet are used to estimate the systematic uncertainty. The unfolding procedure attempts to reconstruct the spectrum of particle-level jets that are matched to detector-level jets using a migration matrix in a model-independent way.  The missed and fake jet distributions, used to correct the matched jet spectra, are however obtained directly from the models.  This introduces a model dependence in the correction procedure.  Because of the lack of $\eta$ segmentation, the contributions from missed and fake jets to the unfolded spectra are relatively large in CASTOR, and the model dependence therefore yields a relatively large uncertainty on the \Pp+Pb~and Pb+\Pp~measurements, and the largest uncertainty on the ratio.
\item \textit{Alignment.} The position of CASTOR is known with limited precision. Events generated by \HIJING are therefore simulated with the CASTOR halves shifted simultaneously inwards and outwards with respect to the beam pipe, according to the alignment uncertainty (about 2\unit{mm} per coordinate in the $x$-$y$ plane). These simulated events are subsequently used to unfold the data. For the detector-level spectra the alignment uncertainty was obtained by reconstructing the spectra with CASTOR simulated at its maximally displaced positions.
\item \textit{Jet identification uncertainty.} Jets are identified to be of hadronic origin before calibration functions are applied. This results in a small fraction of ``true" rejected jets (1 and 0.4\% in \Pp+Pb and Pb+\Pp, respectively).  Since some discrepancy between data and simulation was observed in the distribution of the jet identification variables, the energy spectra are also reconstructed in data without any jet identification, thus treating all reconstructed jets at detector level as being of hadronic origin and letting the unfolding procedure correct for any mismatch between detector and stable-particle level. The impact of this variation is taken as the size of this systematic effect.
\item \textit{Integrated luminosity uncertainty.} The integrated luminosity is known with a precision of 3.4 and 3.6\%~\cite{LumiMeasurement} for \Pp+Pb and Pb+\Pp, respectively.  These uncertainties are treated as independent for the ratio \Pp+Pb/Pb+\Pp.
\end{itemize}

The total systematic uncertainty is obtained by adding the individual sources of uncertainty in quadrature. If a variation in a particular systematic effect results only in a variation of the spectrum in one direction, instead of both, the contribution to the uncertainty is symmetrized. For the leading uncertainty, no symmetrization was needed for the \Pp+Pb and Pb+\Pp\ spectrum. For the subleading sources of uncertainty, for a minor number of bins error symmetrization was needed. For the jet identification uncertainty, which is the least significant source of uncertainty, only one shift was performed and thus the errors are symmetrized for all bins.
The individual contributions and total systematic uncertainty are given in Table~\ref{tab:SystematicUncerttaintiespA} for the \Pp+Pb, Pb+\Pp, and for the ratio of the \Pp+Pb to Pb+\Pp\ cross sections, for their highest and lowest common energy bin. It can be seen that the energy scale uncertainty is dominant for the \Pp+Pb and Pb+\Pp\ spectra at high jet energies, while the model dependence dominates for the ratio of \Pp+Pb to Pb+\Pp\ cross sections.
\begin{table}[htbp]
\centering
\topcaption{The contribution in percentage (\%) of various sources of systematic uncertainty in the highest and lowest common energy bins for the \Pp+Pb, Pb+\Pp, and \Pp+Pb/Pb+\Pp\ spectra.}
\begin{tabular}{ccccccc}
\hline
\multicolumn{1}{c}{} & \multicolumn{2}{c}{\Pp+Pb} &  \multicolumn{2}{c}{Pb+\Pp} & \multicolumn{2}{c}{\Pp+Pb/Pb+\Pp}  \\[\cmsTabSkip]
Energy bin [{\TeVns}] & 0.6
    & 2.5
& 0.6 &2.5
&0.6 &2.5
\\[\cmsTabSkip]\hline\T\B
Energy scale & $^{+2}_{-2}$
& $^{+150}_{-71}$ & $^{+1}_{-2}$
& $^{+120}_{-78}$ & $^{+1}_{-2}$
& $^{+35}_{-35}$    \\[\cmsTabSkip]
Model dependence & $^{+18}_{-18}$
& $^{+41}_{-41}$  & $^{+4}_{-4}$
& $^{+60}_{-60}$ & $^{+1}_{-17}$
& $^{+47}_{-47}$  \\[\cmsTabSkip]
Alignment & $^{+4}_{-4}$
& $^{+34}_{-34}$  & $^{+10}_{-10}$
& $^{+33}_{-33}$ & $^{+14}_{-3}$
& $^{+34}_{-6}$     \\[\cmsTabSkip]
Jet identification & $^{+2}_{-2}$
& $^{+24}_{-24}$  & $^{+2}_{-2}$
&$^{<1}_{<1}$ &$^{<1}_{<1}$
& $^{+25}_{-25}$  \\[\cmsTabSkip]
Total & $^{+19}_{-19}$
& $^{+160}_{-92}$  & $^{+11}_{-11}$
& $^{+140}_{-100}$ & $^{+27}_{-26}$
& $^{+77}_{-54}$  \\[\cmsTabSkip]\hline
\end{tabular}
\label{tab:SystematicUncerttaintiespA}
\end{table}

\section{Results}\label{sec:results}
\begin{figure}[hbtp]
\centering
\includegraphics[width=0.49\textwidth]{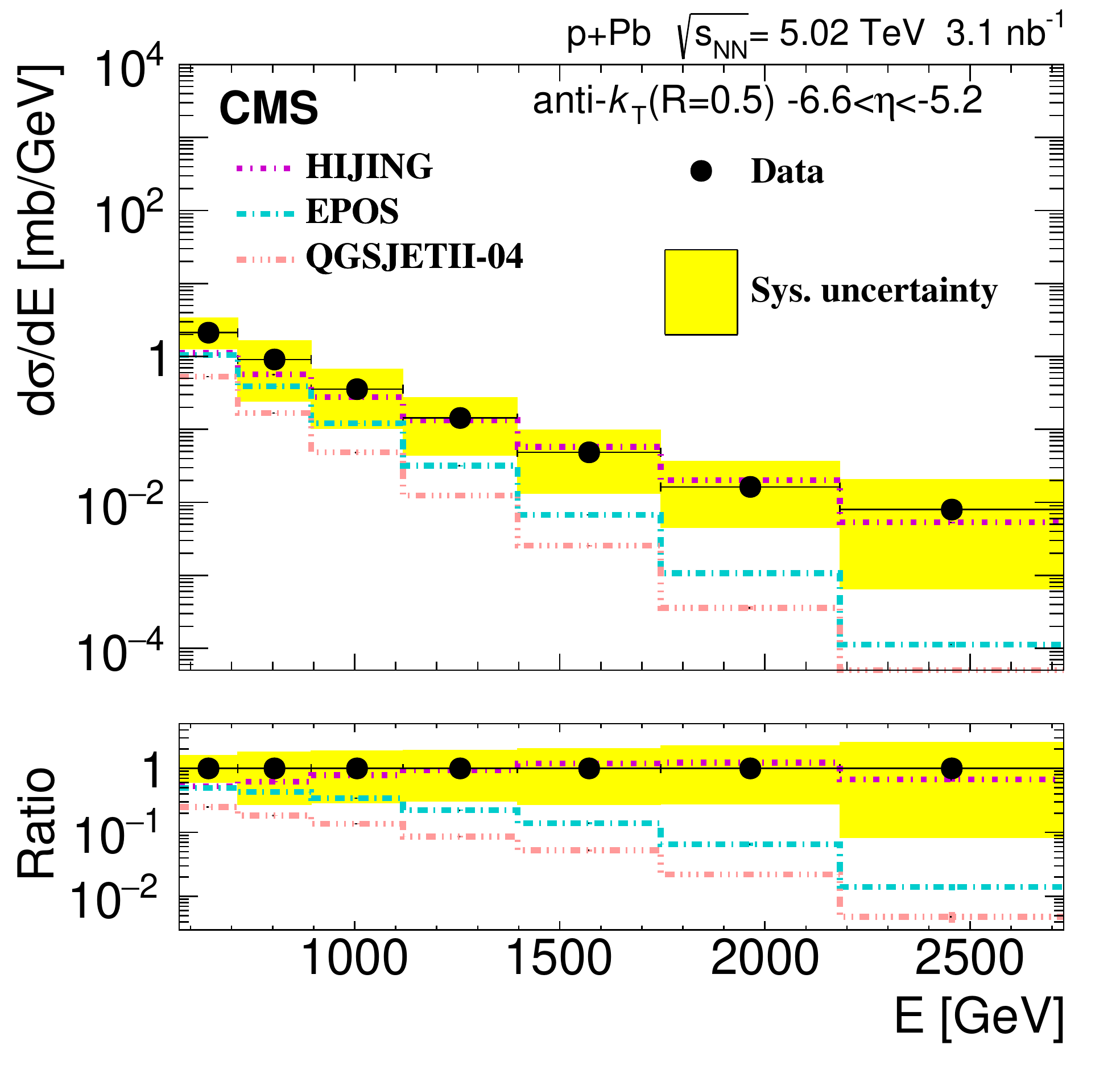}
\includegraphics[width=0.49\textwidth]{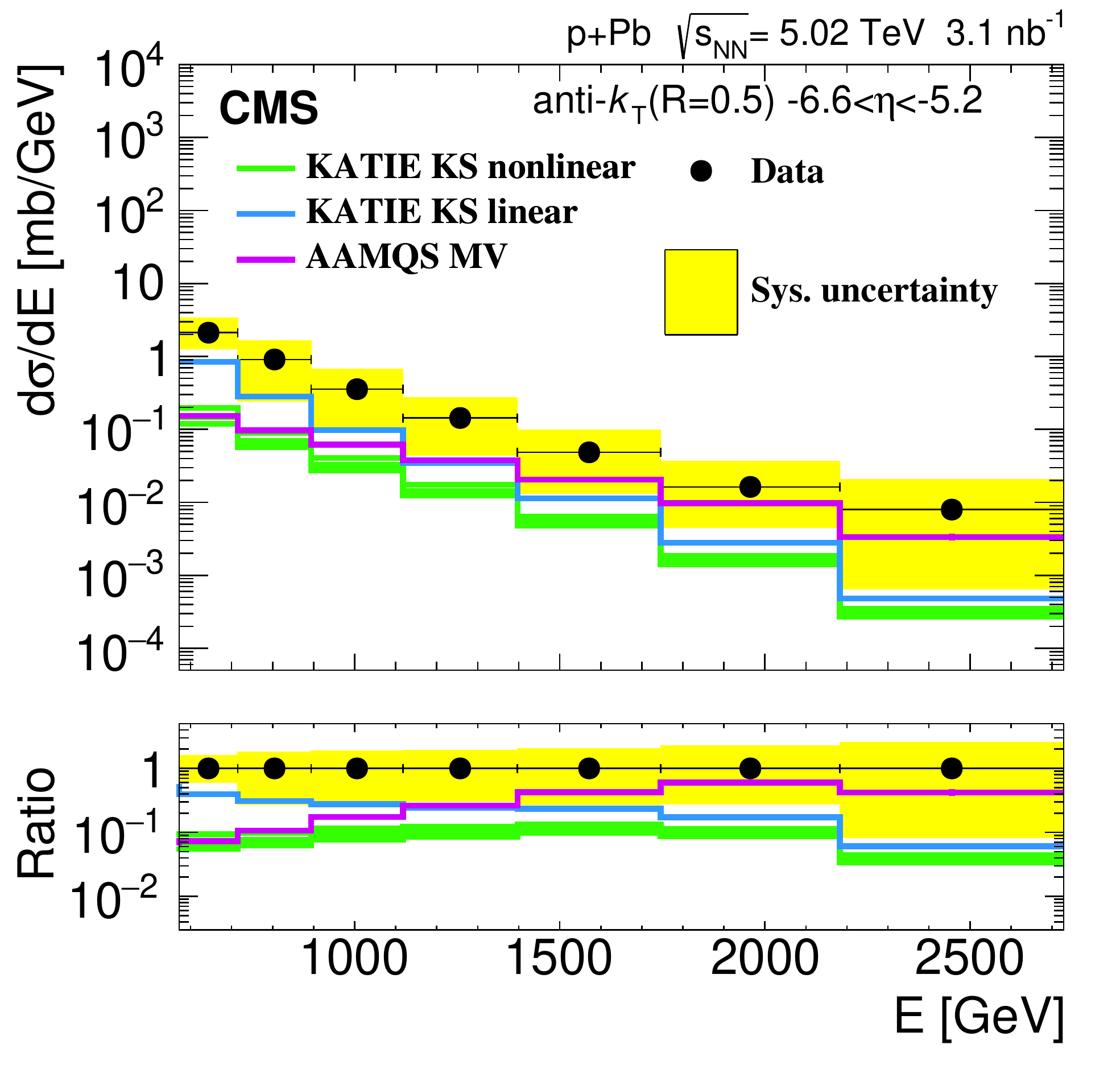}
\caption{
Stable-particle-level differential jet cross section as a function of jet energy measured in \Pp+Pb
collisions at 5.02\TeV, compared to the \EPOS, \HIJING, and \QGSJet (left), and \katie and
\AAMQS (right) predictions. The band associated with the nonlinear \katie curve accounts for
the 50--100\% variation of the strength of the parton saturation effects in this model.}
\label{Plot:pPb}
\end{figure}

\begin{figure}[hbtp]
\centering
\includegraphics[width=0.49\textwidth]{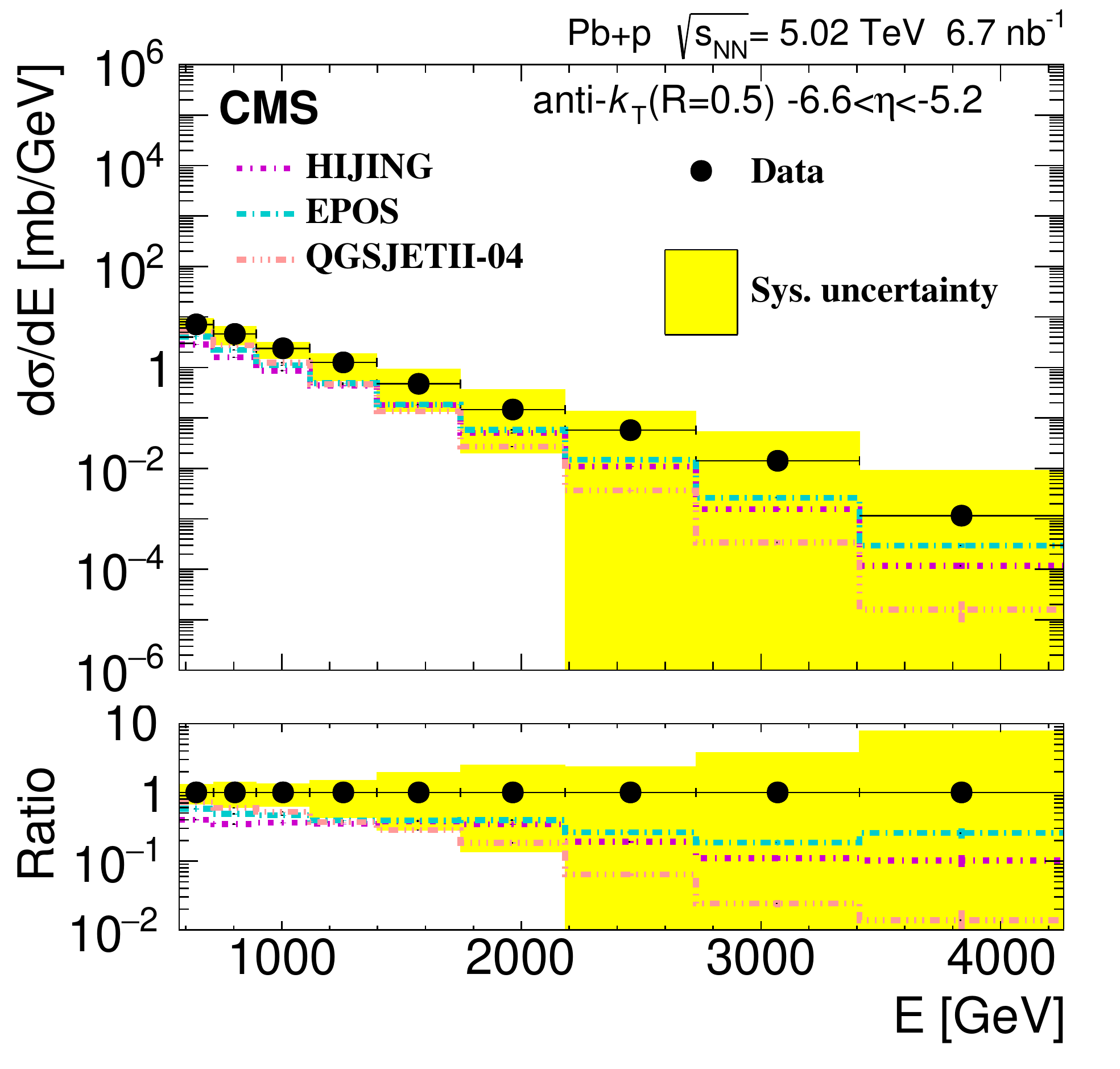}
\includegraphics[width=0.49\textwidth]{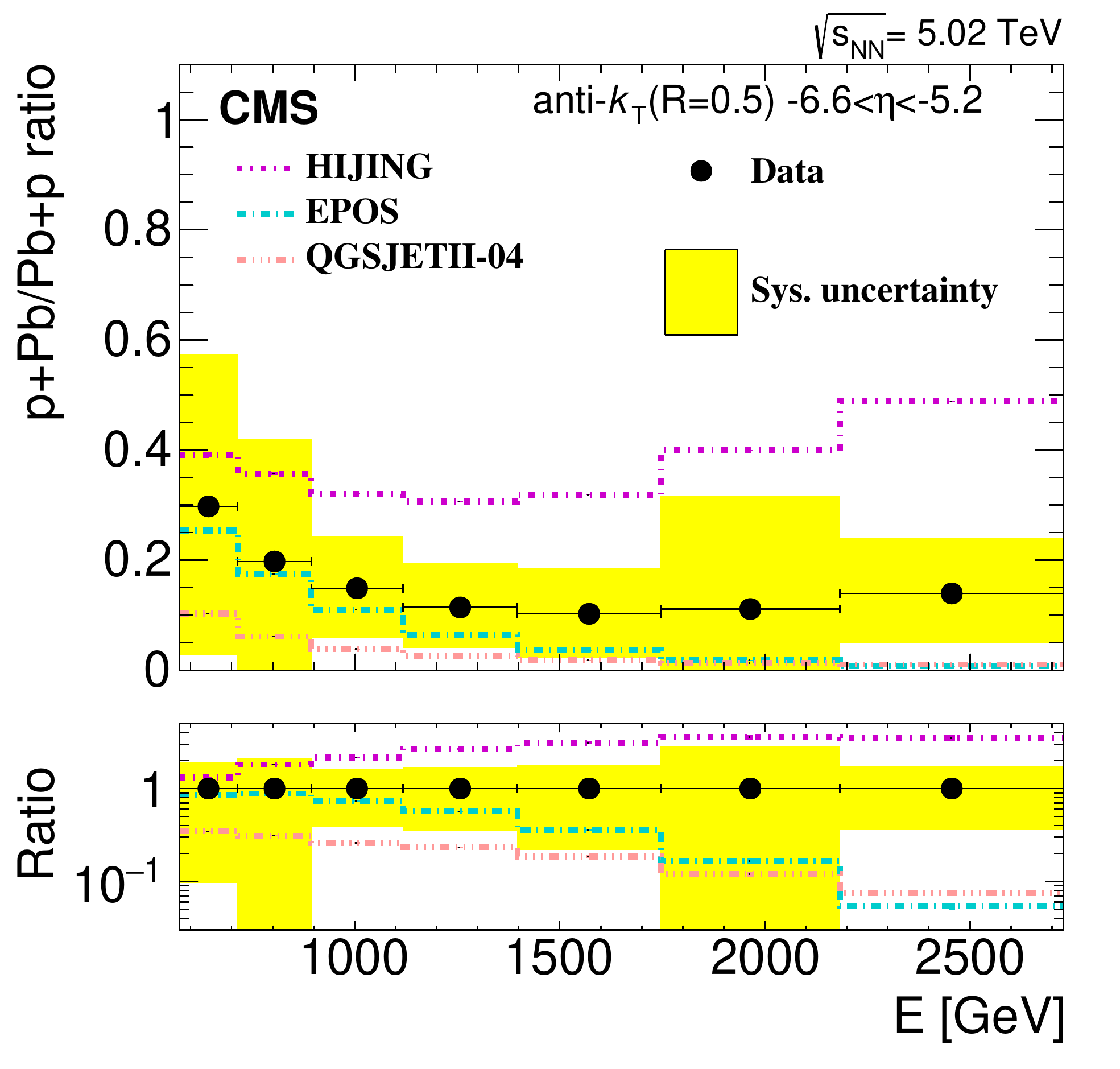}
\caption{Stable-particle-level differential jet cross section as a function of jet energy in proton-lead collisions at 5.02\TeV. The Pb+\Pp\ measurement is depicted left, and the ratio of the \Pp+Pb to Pb+\Pp\ cross sections is displayed right. The data are compared to model predictions from \EPOS, \HIJING, and \QGSJet.}
\label{Plot:PbpAndRatio}
\end{figure}

The unfolded differential cross sections for inclusive forward jet production as a function of jet energy are shown in Fig.~\ref{Plot:pPb} for \Pp+Pb and Fig.~\ref{Plot:PbpAndRatio} (left) for Pb+\Pp\ data.  The ratio of the cross sections in \Pp+Pb to Pb+\Pp\ data is shown in Fig.~\ref{Plot:PbpAndRatio} (right).
The latter distribution would be in principle expected to be directly sensitive to parton saturation effects, independently of any particular data-model comparison. Indeed, whereas jets in the \Pp+Pb data probe the ion parton density at low values of $x$ and are therefore sensitive to possible enhanced saturation effects in nuclei, the Pb+\Pp\ cross section would be sensitive to the low-$x$ content of the proton, which is expected to be less affected by saturation. In principle, strong parton saturation effects would lead to depletion of the cross section ratio
towards low jet energy. However, because the $\eta$ selection is made in the laboratory frame ($-6.6<\eta<-5.2$), and not in the center-of-mass frame, different center-of-mass $\eta$ ranges are being compared (the Lorentz-boost between the \Pp+Pb and Pb+\Pp{} center-of-mass frames results in an $\eta$ shift of $\Delta \eta = 0.87$). The ratio of spectra in Fig.~\ref{Plot:PbpAndRatio} (right) is thus sensitive to proton and lead parton distributions evaluated at
different $x$, thereby complicating the interpretation of the data. Applying a correction for the effect of this boost turns out to be highly model dependent.  The ion debris that contributes to the Pb+\Pp{} cross section further obscures the analysis of this ratio measurement.

In Fig.~\ref{Plot:pPb} (left) the \Pp+Pb data are compared to the predictions of the \EPOS, \HIJING, and \QGSJet models. The predictions of the \EPOS and \QGSJet model differ by more than two orders of magnitude at $E=2.5\TeV$. The \HIJING model describes the measured distributions best, but this does not automatically imply
that its underlying dynamics, DGLAP-based nuclear PDFs with shadowing, is favored by the data, given the many other model ingredients incorporated in this MC generator that affect the results. The \EPOS and \QGSJet models both yield an energy spectrum that is too soft and underestimate the data at high energy.
As mentioned in Section~\ref{sec:generators}, \EPOS suffers from a mismodeling of the absolute normalization of hard scatterings in proton-nucleus collisions that has been improved in more recent versions of the event generator~\cite{Pierog:2015yzd}.

Predictions of the \katie model are compared to data in Fig.~\ref{Plot:pPb} (right).  The two predictions, labeled ``KS nonlinear" and ``KS linear", demonstrate the effect of saturation on the jet cross section, with the KS nonlinear curve being comparatively depleted by an order of magnitude at low jet energy. The band displayed for the nonlinear curve (with gluon saturation) reflects the effect of varying the strength of the saturation term between 50 and 100\%.  Even though the linear approach (without gluon saturation) is closer to the data in terms of normalization, the slope of the nonlinear distribution appears to be a bit harder and more consistent with the shape of the spectrum observed in data.

The \Pp+Pb data are also compared to the predictions from the \AAMQS approach in Fig.~\ref{Plot:pPb} (right).  Even though the approaches used by \katie and \AAMQS are both based on hybrid factorization, the predictions are quite different. This may be traced back to the use of offshell matrix elements in the case of \katie, whereas \AAMQS uses onshell scattering amplitudes derived from the color glass condensate framework. From the comparison to data, it appears that the \AAMQS cross section at low energy is too low, which could be explained by too strong a depletion caused by saturation, while the model describes the data well for high jet energies.

Overall, none of these saturation models can explain all the features observed in the data, and the disagreement between the data and the \katie and \AAMQS saturation models is the largest in the region where nonlinear effects are expected to be the strongest.

Data obtained with the reversed beam configuration are compared to \EPOS, \HIJING, and \QGSJet in Fig.~\ref{Plot:PbpAndRatio} (left). This is the region with significant contributions from ion remnants. The \EPOS and \HIJING models describe the data reasonably well, but are too low in normalization. The \QGSJet model again yields a spectrum that is too soft.

Finally, \EPOS, \HIJING, and \QGSJet predictions are also compared to the ratio of cross sections in \Pp+Pb and Pb+\Pp{} data in Fig.~\ref{Plot:PbpAndRatio} (right), where substantial uncertainty cancelation occurs. None of the models are able to describe the data over the whole range. The \HIJING model describes the shape of the data best, but fails to describe its magnitude because it underestimates the normalization of the Pb+\Pp{} spectrum. The \EPOS model describes the lower energy part of the ratio spectrum well, but fails to describe the shape at high energies because of problems describing the hard component of the jet cross section.
Finally, \QGSJet underestimates both the shape and normalization of the ratio, which can also be attributed to the poor description of the \Pp+Pb spectrum.

The data-model discrepancies described above appear more pronounced in the detector-level distributions plotted in Fig.~\ref{plot:RecoLevelJetESpectrapAAp}, as these latter distributions do not have an associated modeling uncertainty.

\section{Summary}
\label{sec:summary}

Measurements of the differential inclusive forward jet cross sections in proton-lead collisions at $\sqrtsNN=5.02\TeV$ have been presented. The measurements are performed in the laboratory pseudorapidity range $-6.6<\eta<-5.2$, and as a function of jet energy. Collisions with either the incoming proton (\Pp+Pb) or the incoming ion (Pb+\Pp) directed towards the negative $\eta$ hemisphere are studied. The jet cross sections are unfolded to stable-particle level cross sections with $\pt\gtrsim3\GeV$ and compared to predictions from various Monte Carlo event generators.
The cross section ratio for \Pp+Pb to Pb+\Pp\ data as a function of jet energy has also been measured, and exhibits a much smaller systematic uncertainty than the individual spectra.

The so-far unexplored kinematic phase space covered by this measurement is sensitive to the parton densities and their evolution at low fractional momenta. Models incorporating various implementations of gluon saturation have been confronted with data. No model is, however, currently able to describe all aspects of the data.

\begin{acknowledgments}
We congratulate our colleagues in the CERN accelerator departments for the excellent performance of the LHC and thank the technical and administrative staffs at CERN and at other CMS institutes for their contributions to the success of the CMS effort. In addition, we gratefully acknowledge the computing centers and personnel of the Worldwide LHC Computing Grid for delivering so effectively the computing infrastructure essential to our analyses. Finally, we acknowledge the enduring support for the construction and operation of the LHC and the CMS detector provided by the following funding agencies: BMWFW and FWF (Austria); FNRS and FWO (Belgium); CNPq, CAPES, FAPERJ, and FAPESP (Brazil); MES (Bulgaria); CERN; CAS, MoST, and NSFC (China); COLCIENCIAS (Colombia); MSES and CSF (Croatia); RPF (Cyprus); SENESCYT (Ecuador); MoER, ERC IUT, and ERDF (Estonia); Academy of Finland, MEC, and HIP (Finland); CEA and CNRS/IN2P3 (France); BMBF, DFG, and HGF (Germany); GSRT (Greece); NKFIA (Hungary); DAE and DST (India); IPM (Iran); SFI (Ireland); INFN (Italy); MSIP and NRF (Republic of Korea); LAS (Lithuania); MOE and UM (Malaysia); BUAP, CINVESTAV, CONACYT, LNS, SEP, and UASLP-FAI (Mexico); MBIE (New Zealand); PAEC (Pakistan); MSHE and NSC (Poland); FCT (Portugal); JINR (Dubna); MON, RosAtom, RAS and RFBR (Russia); MESTD (Serbia); SEIDI, CPAN, PCTI and FEDER (Spain); Swiss Funding Agencies (Switzerland); MST (Taipei); ThEPCenter, IPST, STAR, and NSTDA (Thailand); TUBITAK and TAEK (Turkey); NASU and SFFR (Ukraine); STFC (United Kingdom); DOE and NSF (USA).

\hyphenation{Rachada-pisek} Individuals have received support from the Marie-Curie program and the European Research Council and Horizon 2020 Grant, contract No. 675440 (European Union); the Leventis Foundation; the A. P. Sloan Foundation; the Alexander von Humboldt Foundation; the Belgian Federal Science Policy Office; the Fonds pour la Formation \`a la Recherche dans l'Industrie et dans l'Agriculture (FRIA-Belgium); the Agentschap voor Innovatie door Wetenschap en Technologie (IWT-Belgium); the F.R.S.-FNRS and FWO (Belgium) under the ``Excellence of Science - EOS" - be.h project n. 30820817; the Ministry of Education, Youth and Sports (MEYS) of the Czech Republic; the Lend\"ulet (``Momentum") Program and the J\'anos Bolyai Research Scholarship of the Hungarian Academy of Sciences, the New National Excellence Program \'UNKP, the NKFIA research grants 123842, 123959, 124845, 124850 and 125105 (Hungary); the Council of Science and Industrial Research, India; the HOMING PLUS program of the Foundation for Polish Science, cofinanced from European Union, Regional Development Fund, the Mobility Plus program of the Ministry of Science and Higher Education, the National Science Center (Poland), contracts Harmonia 2014/14/M/ST2/00428, Opus 2014/13/B/ST2/02543, 2014/15/B/ST2/03998, and 2015/19/B/ST2/02861, Sonata-bis 2012/07/E/ST2/01406; the National Priorities Research Program by Qatar National Research Fund; the Programa Estatal de Fomento de la Investigaci{\'o}n Cient{\'i}fica y T{\'e}cnica de Excelencia Mar\'{\i}a de Maeztu, grant MDM-2015-0509 and the Programa Severo Ochoa del Principado de Asturias; the Thalis and Aristeia programmes cofinanced by EU-ESF and the Greek NSRF; the Rachadapisek Sompot Fund for Postdoctoral Fellowship, Chulalongkorn University and the Chulalongkorn Academic into Its 2nd Century Project Advancement Project (Thailand); the Welch Foundation, contract C-1845; and the Weston Havens Foundation (USA).
\end{acknowledgments}
\bibliography{auto_generated}
\cleardoublepage \appendix\section{The CMS Collaboration \label{app:collab}}\begin{sloppypar}\hyphenpenalty=5000\widowpenalty=500\clubpenalty=5000\vskip\cmsinstskip
\textbf{Yerevan Physics Institute, Yerevan, Armenia}\\*[0pt]
A.M.~Sirunyan, A.~Tumasyan
\vskip\cmsinstskip
\textbf{Institut f\"{u}r Hochenergiephysik, Wien, Austria}\\*[0pt]
W.~Adam, F.~Ambrogi, E.~Asilar, T.~Bergauer, J.~Brandstetter, E.~Brondolin, M.~Dragicevic, J.~Er\"{o}, A.~Escalante~Del~Valle, M.~Flechl, V.M.~Ghete, J.~Hrubec, M.~Jeitler\cmsAuthorMark{1}, N.~Krammer, I.~Kr\"{a}tschmer, D.~Liko, T.~Madlener, I.~Mikulec, N.~Rad, H.~Rohringer, J.~Schieck\cmsAuthorMark{1}, R.~Sch\"{o}fbeck, M.~Spanring, D.~Spitzbart, A.~Taurok, W.~Waltenberger, J.~Wittmann, C.-E.~Wulz\cmsAuthorMark{1}, M.~Zarucki
% ...
\vskip\cmsinstskip
% ...
1:  Also at Vienna University of Technology, Vienna, Austria\\
% ...
\end{sloppypar}
\end{document}